\documentclass[sigconf]{acmart}
\AtBeginDocument{%
  \providecommand\BibTeX{{%
    \normalfont B\kern-0.5em{\scshape i\kern-0.25em b}\kern-0.8em\TeX}}}
\newcommand{\ie}{\textit{i}.\textit{e}.}
\newcommand{\eg}{\textit{e}.\textit{g}.}
\newcommand{\etal}{\textit{et al}.}

\usepackage{times}
\usepackage{epsfig}
\usepackage{graphicx}
\usepackage{amsmath}
\usepackage{pifont}
\usepackage{arydshln}
\usepackage{diagbox}
\usepackage{booktabs}
\usepackage{multirow}

\newcommand{\cmark}{\ding{51}}

\setcopyright{acmcopyright}
\copyrightyear{2023}
\acmYear{2023}
\acmDOI{10.1145/3581783.3613457} 
\acmConference[MM'23]{Make sure to enter the correct conference title from your rights confirmation emai}{October 29 -- November 3, 2023}{Ottawa, Canada}
\acmBooktitle{Proceedings of the 31th ACM International Conference on Multimedia (MM'23), October 29 -- November 3, 2023. Ottawa, Canada} 
\acmPrice{15.00}
\begin{document}
\title{Screen-based 3D Subjective Experiment Software}
\author{Songlin Fan}
\email{slfan@pku.edu.cn}
\affiliation{%
\institution{Peking University Shenzhen Graduate
School}
\institution{Pengcheng Laboratory}
\country{China}}



\author{Wei Gao}
\authornote{Corresponding author. This work was supported by Natural Science Foundation of China (62271013, 62031013), Shenzhen Fundamental Research Program (GXWD20201231165807007-20200806163656003), Shenzhen Science and Technology Plan Basic Research Project (JCYJ20190808161805519), and was sponsored by CAAI-Huawei MindSpore Open Fund (CAAIXSJLJJ-2022-002C).}
\email{gaowei262@pku.edu.cn}
\affiliation{%
  \institution{Peking University Shenzhen Graduate School}
  \country{China}
}

\begin{abstract}
Recently, widespread 3D graphics (\eg, point clouds and meshes) have drawn considerable efforts from academia and industry to assess their perceptual quality by conducting subjective experiments. However, lacking a handy software for 3D subjective experiments complicates the construction of 3D graphics quality assessment datasets, thus hindering the prosperity of relevant fields. In this paper, we develop a powerful platform with which users can flexibly design their 3D subjective methodologies and build high-quality datasets, easing a broad spectrum of 3D graphics subjective quality study. To accurately illustrate the perceptual quality differences of 3D stimuli, our software can simultaneously render the source stimulus and impaired stimulus and allows both stimuli to respond synchronously to viewer interactions. Compared with amateur 3D visualization tool-based or image/video rendering-based schemes, our approach embodies typical 3D applications while minimizing cognitive overload during subjective experiments. We organized a subjective experiment involving 40 participants to verify the validity of the proposed software. Experimental analyses demonstrate that subjective tests on our software can produce reasonable subjective quality scores of 3D models. All resources in this paper can be found at \textcolor{red}{\url{https://openi.pcl.ac.cn/OpenDatasets/3DQA}}.
\end{abstract}

\ccsdesc[500]{Experience~Interactions and Quality of Experience}
\keywords{Subjective Experiment Software, Quality Assessment, 3D Graphics, Point Cloud, Mesh. }
\maketitle
\section{Introduction}\label{sec:1}
\begin{figure}[t]
\centering
\includegraphics[width=\linewidth]{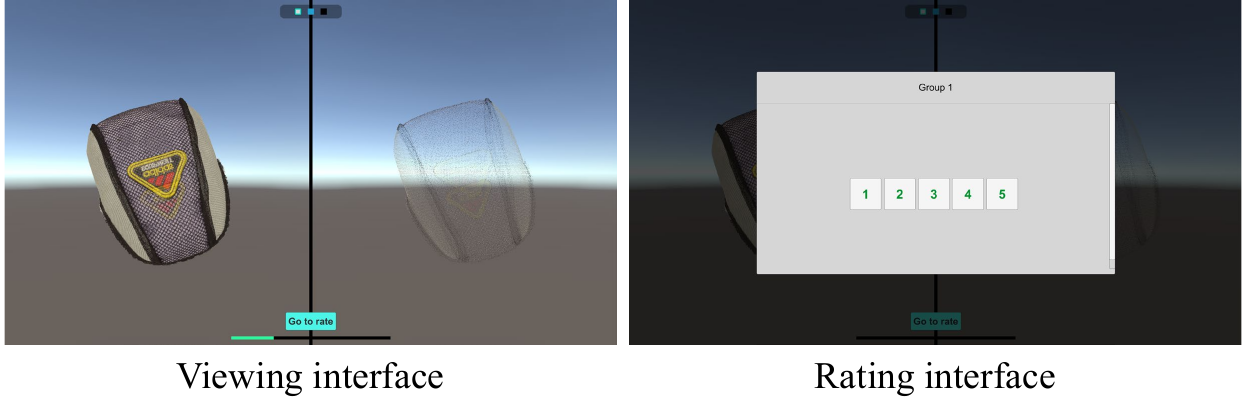}
\caption{Windows of our 3D subjective experiment software.}
\label{fig1}
\end{figure}

In the past decade, prevalent 3D acquisition and display devices have facilitated various 3D graphics applications~\cite{fan2022salient}, triggering an era of 3D data explosion. The size and complexity of 3D data are increasingly exceeding the processing capabilities of existing computing devices, especially portable devices. Therefore, simplification and compression operations are often applied to the source 3D data to reduce computing costs, which inevitably degrade their perceptual quality. Quantitatively assessing the perceived quality change of 3D data is critical to improving the user experience and the performance of relevant compression/enhancement algorithms.

Methods for measuring the perceptual quality of 3D graphics focus mainly on point clouds and meshes, including subjective tests and objective metrics. Subjective tests mean recruiting several participants to conduct subjective experiments to assess the quality of 3D graphics. Considering that humans often play the role of ultimate viewers, subjective tests are the most reliable approach to evaluating 3D graphics quality, which also provide the foundation for objective metric designs. However, the lack of an appropriate subjective experiment software makes it highly inconvenient to carry out 3D content subjective tests. Existing representative screen-based methods can be classified into two genres, \ie, image/video rendering-based methods~\cite{nehme2022textured,liu2022perceptual} and 3D visualization tool-based methods~\cite{yang2020predicting,liu2023point}. The former genre first projects 3D graphics into 2D images and evaluates the quality of projections from controlled viewpoints, sacrificing viewing freedom. The latter genre leverages existing visualization tools (such as CloudCompare\footnote{\url{https://www.danielgm.net/cc/}}) to display the source stimulus and impaired stimulus individually and operate subjective experiments manually, which may increase cognitive overload~\cite{nehme2022textured} and result in judgment fluctuations. Consequently, the community urgently demands a well-designed 3D subjective experiment platform.

In this work, we present a software tailored for 3D graphics subjective experiments (see Figure~\ref{fig1}), on which subjective tests can be flexibly designed and conveniently implemented. The proposed software supports the direct rendering of multiple 3D content formats (such as point clouds and meshes) and potentially catalyzes the development of relevant fields. The size of rendered 3D models can be set freely according to the subjective methodology designs. Our software can simultaneously render the source stimulus and impaired stimulus for experiment participants to compare their visual differences from free viewpoints. The advantages of simultaneous rendering for 3D content over alternatives have been demonstrated by Alexiou \etal~\cite{alexiou2017performance}. To reduce cognitive overload and judgment fluctuations~\cite{nehme2022textured}, we realize the synchronous response of stimuli to viewer interactions in our software, facilitating the viewers in perceiving the visual differences between two stimuli. Additionally, our proposed software also supports other vital features, such as flexible display modes, timer, breakpoint startup, and background switching, to name a few. These featured settings enable the software to meet the diverse experimental needs of users.

To illustrate the reasonableness and validity of our scheme, we collect five high-quality point clouds as the source stimuli and generate 40 impaired models with varying compression distortions. We then conduct a cross-validation experiment involving 40 participants using our software. Experimental results indicate that subjective tests through our software can yield consistent subjective scores reasonably reflecting the perceptual quality of 3D graphics. It is worth noting that \textbf{our software is accepted by the Audio Video coding Standard (AVS)~\cite{M7666} as a reference software and will continue to be updated. The resulting dataset is also publicly available.}


\section{Features and Usage}

\begin{figure}[t]
\centering
\includegraphics[width=0.80\linewidth]{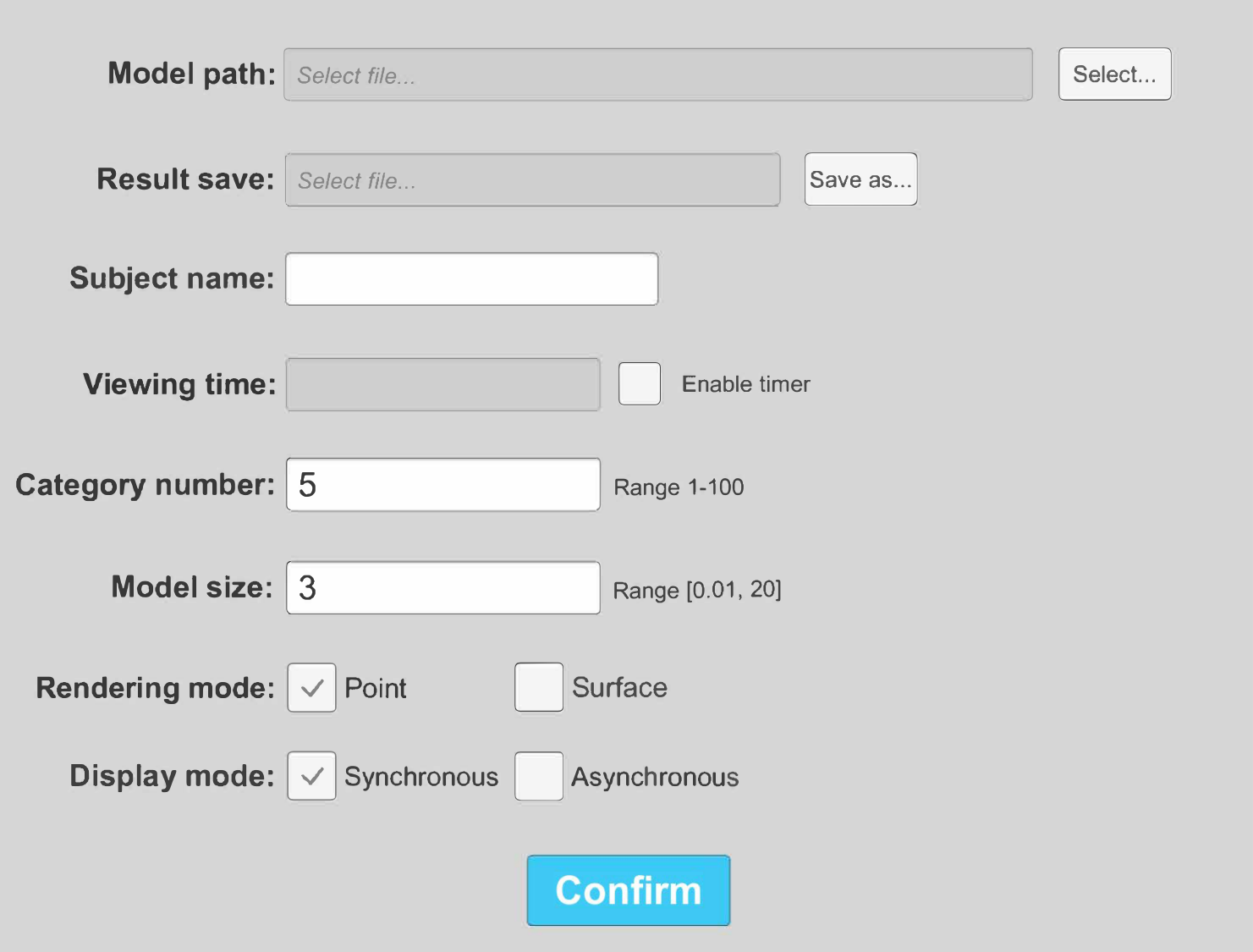}
\caption{Software setup interface with flexible options.}
\label{fig2}
\end{figure}

The proposed subjective experiment software is developed on Unity{\footnote{\url{https://unity.com/}}} with user-friendly features that simplify the setup and execution of subjective experiments for 3D graphics. As shown in Figure~\ref{fig2}, researchers can customize their experimental parameters to achieve relevant methodology designs. This section will first detail several important features of our software for 3D subjective experiments and then introduce how to conduct experiments on this software. 

\subsection{Software Features}

\noindent\textbf{Direct 3D rendering.} The proposed software is tailored for the subjective quality study of 3D content, where 3D models can be rendered as points or surfaces directly. Therefore, our method introduces no additional data conversions compared to previous image/video rendering-based methods. The data conversions of previous methods not only increase computational overhead but may alter the original perceived quality of 3D models, ultimately impacting the accuracy of subjective experiments. Furthermore, experiment participants using our software can watch rendered 3D models from any viewpoint through rotation operations, known as interactive subjective experiments, which fully mimic typical 3D applications.

\noindent\textbf{Multiple 3D formats.}  The exploration of 3D graphics quality assessment is still in its infancy, and the measurement of perceptual quality for many 3D data formats still needs improvement. We develop our subjective experiment platform supporting multiple data formats. Specifically, our software is capable of directly taking various standard 3D content formats (such as PLY and OBJ files) as inputs, which are widely adopted for point clouds and meshes in numerous research and application fields. Therefore, our software has significant implications for accelerating the accurate modeling of perceptual quality for point clouds and meshes. 

\noindent\textbf{Flexible display modes.} Interactive subjective experiments often result in cognitive overload, altering viewer judgments, particularly when paired stimuli are displayed independently. Our software can simultaneously display the source stimulus and impaired stimulus, effectively mitigating cognitive overload. It allows participants to view the stimuli side by side, facilitating direct comparisons and evaluations. This approach lets participants easily perceive the visual differences between the paired stimuli, leading to more consistent subjective quality judgments of corrupted stimuli. While the simultaneous display mode is recommended in our software, it also supports other display modes to accommodate customized needs.

\noindent\textbf{Timer.} Timers are frequently adopted to provide accurate and consistent timing measurements in subjective experiments. They ensure that participants have a predefined time limit to evaluate each 3D model, thereby eliminating the possibility of biased or rushed evaluations. We integrate an optional timer into the proposed software to facilitate the efficient management of subjective experiments. In addition, the timer allows participants to continue viewing even after the predefined time ends, in accordance with the recommendations of the AVS standard~\cite{M7666}.

\noindent\textbf{Breakpoint startup.} As it is difficult for participants to maintain enduring concentration, the subjective experiment process often consists of several time slices interspersed with rest periods. To further convenience researchers to implement large-scale subjective experiments, we make our software have the function of breakpoint startup. This feature allows users to pause the experiment process at any point and resume it later, exhibiting significant flexibility and convenience. Moreover, the breakpoint startup function enables researchers to handle interruptions or prioritize more important tasks without losing progress.

\noindent\textbf{Background switching.} Considering the varying color distributions of 3D models, our software allows users to change the visual backdrop during subjective experiments, providing diverse viewing environments. Furthermore, the scene switching feature aids in detecting potential distortions or artifacts that may be more visible against specific backgrounds, thereby improving the applicability of our software to 3D data with varying color attributes.

\begin{figure}[t]
\centering
\includegraphics[width=0.95\linewidth]{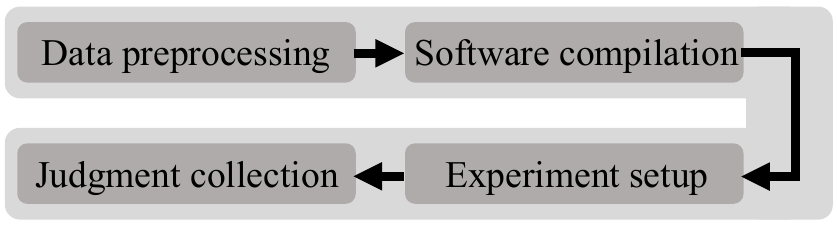}
\caption{Usage procedures of our software.}
\label{fig3}
\end{figure}

\subsection{Usage Procedures} 
As illustrated in Figure~\ref{fig3}, the whole usage procedures of our software consist of four standard steps, \ie, data preprocessing, software compilation, experiment setup, and judgment collection. Data preprocessing involves normalizing the size of models to be viewed, ensuring that all models have the same base size, and preventing rendering issues caused by model size differences. The processed models are then transferred to our software project. Software compilation refers to packaging the project into executable desktop software that can start subjective experiments with a single click. Once the compiled software is launched, the experiment setup involves configuring the software settings according to specific methodology designs. Users can customize experiment parameters such as model display order, result save path, participant name, viewing time, distortion level number, model size, rendering mode, and display mode. Finally, judgment collection represents the process of conducting subjective experiments, where users assess the visual quality of 3D models following predefined protocols.


\section{Experiments}
Our software is proposed to address the lack of a well-designed platform for 3D graphics subjective experiments, with features specifically tailored for 3D content. Subjective tests adopting our software can generate accurate and consistent quality judgments for 3D graphics. To experimentally validate the effectiveness of the proposed software, we conduct a cross-validation experiment to analyze the performance of our platform.

\subsection{Stimulus Collection}

\begin{figure}[t]
\centering
\includegraphics[width=0.7\linewidth]{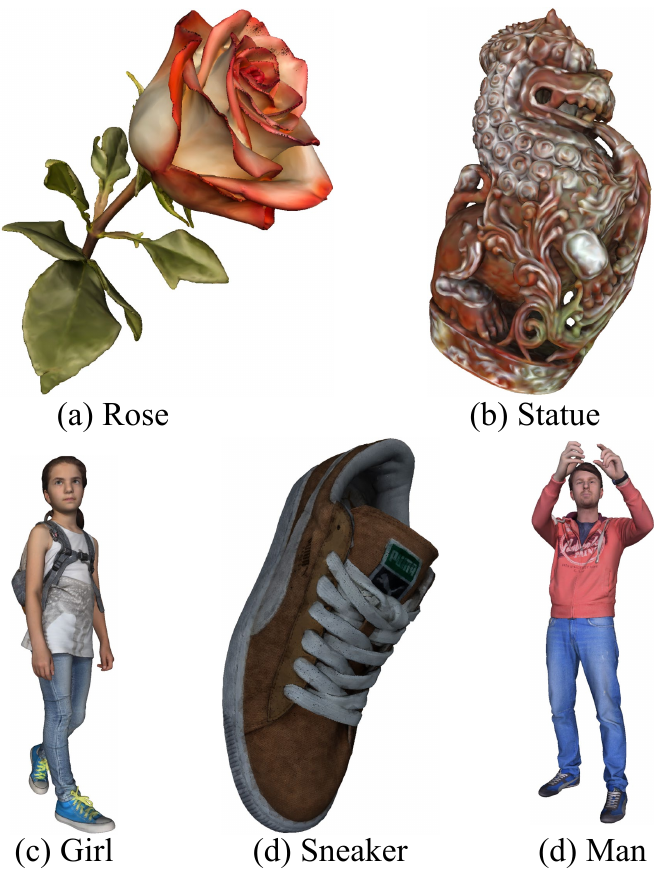}
\caption{Our collected source models including (a) Rose, (b) Statue, (c) Girl, (d) Sneaker, and (d) Man.}
\label{fig4}
\end{figure}

\noindent\textbf{Source stimuli.} 
We manually collect over twenty high-quality point cloud models from the Internet, all provided in a standard PLY format. The visual quality of each model is carefully checked, and we select the five highest-quality models with different content as the source stimuli. Figure~\ref{fig4} illustrates the snapshots of the selected models, namely “Rose,” “Statue,” “Girl,” “Sneaker,” and “Man.” Each model consists of more than 600,000 points with associated color and normal attributes, representing a distinct object category found in daily life. We strive to ensure a broad coverage of future usage situations with the selected content. Apart from the object category, the geometric and textural complexity of these 3D models are also considered to ensure a diverse dataset. 

\noindent\textbf{Impaired stimuli.} 
Due to the large size of point clouds, compression operations are often necessary for data storage and transmission. Compression artifacts, one of the most common distortion types, can significantly degrade the quality of point clouds. Therefore, we apply compression algorithms to the collected source stimuli to generate corrupted stimuli with varying levels of visual quality degradation. In particular, we utilize the AVS compression tool, namely Point Cloud Reference Model (PCRM)~\cite{N3445}, to obtain impaired stimuli from the five source stimuli. Since changes in both geometry and attribute information of point clouds may alter their perceptual quality, we consider applying eight combinations of different geometry and attribute compression parameters~\cite{M7666} for generating impaired stimuli with rich and uniform quality variations. The compression parameters we selected are shown in Table~\ref{tab1}, covering a range of low, medium, and high compression efficiency. Consequently, we eventually obtain $5\ (source\ stimuli)\times 8 \ (distortions) = 40$ impaired stimuli for subsequent subjective tests.

\begin{table}[t]
\renewcommand{\tabcolsep}{2.5mm}
\centering
\begin{tabular}{l|cccc}
\toprule
\diagbox{Geometry}{Attribute}&r6 & r3 & r2 &r1 \\ \midrule
r5 & \cmark &\cmark&\cmark& \cmark\\ 
r2 & \cmark &\cmark&&\\ 
r1 & \cmark &&&\cmark  \\
\bottomrule
\end{tabular}
\caption{Selection of compression parameters~\cite{M7666} for generating compressed point cloud samples.}
\label{tab1}
\end{table}	

\subsection{Subjective Tests}
The environment of our subjective tests follows the recommendation of BT.500-13~\cite{rec2012bt}. The display device is a 27-inch monitor with 4K resolution, and the platform is our 3D subjective experiment software. The experimental settings of our software keep defaults with the synchronous display and point rendering modes selected. The timer in our software is enabled, and we recommend a viewing time of around 20 seconds before making quality judgments. The category number of quality judgments is five, represented by from 1 (the worst quality) to 5 (the best quality). 

We recruit 40 subjects aged between 18 and 50 with balanced genders, which are then randomly divided into two equal groups (Group-1 and Group-2) for a cross-validation experiment. Before the formal experiment, each participant learns the purpose and operation of this subjective experiment by watching the teaching video. Subsequently, we will ask every participant to rate the visual quality of all impaired stimuli by comparing them with corresponding source stimuli. While viewing, observers can freely rotate point cloud samples according to their preferences. Furthermore, we also insert two trapping samples that display twice at random for each source point cloud. The trapping samples are adopted to evaluate the judgment accuracy of subjects and identify unreliable subjects. If a participant fails to rate a trapping sample correctly, \ie, ratings for the same sample differ by more than 2, all judgments from this participant will be discarded. 

\subsection{Result Analyses}
Our subjective tests receive 2,000 quality judgments from 40 subjects. Attentive checks find that one participant in Group-1 and two participants in Group-2 are detected as unreliable subjects. We reject all judgments from these unreliable subjects and only retain those from qualified subjects for succeeding analyses.

\begin{figure}[t]
\centering
\includegraphics[width=0.85\linewidth]{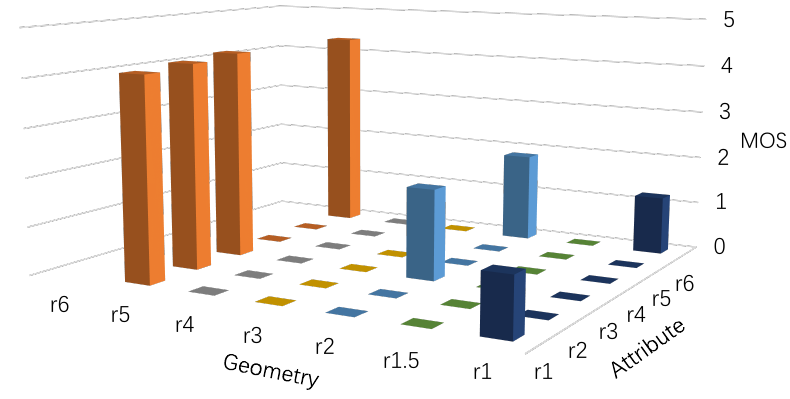}
\caption{MOS distribution of different compression parameters in our subjective experiment.}
\label{fig5}
\end{figure}

\noindent\textbf{Reasonableness verification.}
To observe the reasonableness of quality judgments from our subjective experiment, we compute the Mean Opinion Score (MOS) of each impaired stimulus by averaging the rating scores from two groups of participants. Figure~\ref{fig5} shows the distribution of MOSs with respect to the compression parameters of all impaired stimuli. Note that a smaller ordinal number of the compression parameter indicates a higher compression degree (worse visual quality). We can learn that the MOS correlates with the compression degree visibly. 
Specifically, the increase in the degree of geometry or attribute compression can drop the MOS of 3D models, indicating that the variation of MOSs from our software reasonably reflects the quality change of 3D graphics.

\noindent\textbf{Consistency verification.}
To study the stability of our software in subjective tests, we compute the MOS from two groups of qualified subjects separately and perform a cross-validation analysis. We adopt four widely-used evaluation metrics in the quality assessment field~\cite{9428383} to measure the correlation of MOSs between Group-1 and Group-2, including Spearman Rank Order Correlation Coefficient (SROCC), Pearson Linear Correlation Coefficient (PLCC), Kendall’s Rank Order Correlation Coefficient (KROCC), and Root Mean Squared Error (RMSE). Based on these four metrics, we can obtain convincing evaluation results. The cross-validation results are listed in Table~\ref{tab2}, from which we can learn that the quality judgments for 3D models from two groups of participants exhibit a high degree of agreement. We attribute the consistent judgments to the simultaneous display and synchronous rendering of our software because subjects can easily find the perceptual differences of paired stimuli based on our scheme.

\begin{table}[t]
\renewcommand{\tabcolsep}{1.2mm}
\centering
\begin{tabular}{l|cccc}
\toprule
 Metric &SROCC $\uparrow$ & PLCC$\uparrow$ & KROCC$\uparrow$ &RMSE$\downarrow$ \\ \midrule
Group 1 \textit{vs} Group-2 & 0.9632 &0.9962 &0.8723 &0.0318  \\
\bottomrule
\end{tabular}
\caption{Cross-validation results for two groups of subjects. Note that the results are computed on the normalized MOS. $``\uparrow"/``\downarrow"$ indicates that larger/smaller is better.}
\label{tab2}
\end{table}	
\section{Conclusion}
This paper presents a 3D subjective experiment software, addressing the lack of a convenient platform for assessing the perceptual quality of widespread 3D graphics. The proposed software provides users with various experiment settings to design subjective methodologies and potentially catalyze a broad spectrum of 3D graphics subjective quality study. By simultaneously rendering the source and impaired stimuli and allowing synchronous viewer interactions, our software accurately shows users the visual quality differences of 3D stimuli. To illustrate the effectiveness of our software, we further conduct a cross-validation experiment involving 40 subjects. Experimental results show that subjective tests with our software can obtain convincing and consistent quality judgments of 3D graphics. 

\bibliographystyle{plain}
\bibliography{sample-base}
\end{document}